# Strain tuning of a quantum dot strongly coupled to a photonic crystal cavity

Shuo Sun,[1] Hyochul Kim,[1] Glenn S. Solomon,[2] and Edo Waks[1,a]

[1]Department of Electrical and Computer Engineering, IREAP, and Joint Quantum Institute, University of Maryland, College Park, Maryland 20742, USA

[2]Joint Quantum Institute, National Institute of Standards and Technology, and University of Maryland, Gaithersburg, Maryland 20899, USA

**Abstract:** We demonstrate reversible strain-tuning of a quantum dot strongly coupled to a photonic crystal cavity. We observe an average redshift of 0.45 nm for quantum dots located inside the cavity membrane, achieved with an electric field of 15 kV/cm applied to a piezo-electric actuator. Using this technique, we demonstrate the ability to tune a quantum dot into resonance with a photonic crystal cavity in the strong coupling regime, resulting in a clear anti-crossing. The bare cavity resonance is less sensitive to strain than the quantum dot and shifts by only 0.078 nm at the maximum applied electric field.

Semiconductor quantum dots coupled to photonic crystal cavities offer a robust and scalable platform for studying cavity quantum electrodynamics in a solid state device. The high quality factor and small mode volume of photonic crystal cavities, coupled with the large oscillator strength of quantum dots, enable light-matter interactions in the strong coupling regime.[1,2] This regime is a prerequisite for applications such as cavity reflectivity control,[3] single-photon level optical nonlinearities,[4,5] low photon number all-optical switching,[6-8] and qubit-photon logic operations[9].

---

[a] Author to whom correspondence should be addressed. Electronic mail: edowaks@umd.edu.

The study of quantum dots coupled to photonic crystal cavities often requires the ability to tune the quantum dot exciton energy *in-situ*. Tuning compensates for spectral mismatch between the quantum dot exciton energy and the cavity resonant frequency, and also provides control over the interaction strength between the two systems. The quantum dot exciton energy can be tuned by various means such as changing the sample temperature,[1,3] applying an AC Stark shift,[10] utilizing the quantum confined Stark effect,[11] or applying a magnetic field.[12-14] Temperature tuning and AC Stark effect can broaden the linewidth of the quantum dot through phonon scattering[15] or excitation induced dephasing[16-19]. The quantum confined Stark effect does not suffer from these drawbacks but can quench quantum dot emission at high electric fields due to separation of the electron and hole wavefunctions.[20] Magnetic field tuning can reduce the coupling strength between the quantum dot and the cavity by changing the polarization of the emitted light and by magnetic field induced carrier confinement which lowers the quantum dot oscillator strength.[12]

Strain tuning is an alternate method for tuning quantum dots.[21-33] In this method strain modifies the confinement of electrons and holes, thereby changing their Coulomb interaction strength.[24] Strain tuning can reversibly shift the quantum dot exciton energy without affecting its emission linewidth or intensity,[24] and can achieve a very large tuning range of up to 20 meV[33]. These advantages make strain a promising method for tuning quantum dots coupled to optical cavities. Strain tuning of a single quantum dot strongly coupled to a microdisk cavity has been previously reported.[31] However, applying strain to a photonic crystal cavity can be challenging because these structures are supported on suspended membranes. Recently, new methods to tune the resonances of photonic crystal cavity modes with strain have been demonstrated.[34,35] These techniques open up the possibility for strain tuning quantum dots in a photonic crystal device.

In this letter, we demonstrate reversible *in-situ* strain tuning of a quantum dot strongly coupled to a photonic crystal cavity. We achieve an average strain induced shift of 0.45 nm and demonstrate anti-crossing of a quantum dot and a photonic crystal cavity mode. We show that inside the cavity membrane, quantum dots exhibit large fluctuations in the strain induced shift as compared to those in the bulk, and also shift in the opposite direction. These results indicate that the strain in the membrane region is very different from that of the bulk. We also demonstrate that the cavity mode resonance shifts by an amount that is 5.8 times smaller than the typical quantum dot shift. Thus, the quantum dot can be tuned over an appreciable range with only a small change to the cavity resonance.

We perform measurements on a GaAs photonic crystal cavity coupled to InAs quantum dots. The initial wafer is composed of a 160 nm GaAs membrane with a single layer of InAs quantum dots at its center with a density of 10-50/µm$^2$. The membrane layer is grown on top of a 900 nm

Al$_{0.78}$Ga$_{0.22}$As sacrificial layer. The GaAs is patterned by electron beam lithography, followed by chlorine based dry etching. After dry etching, a selective wet etching removes the sacrificial layer, creating a suspended membrane. The cavity design is based on a three-hole defect in a triangular photonic crystal where the inner three holes adjacent to the defect are shifted to optimize the quality factor.[36] Figure 1(a) shows a scanning electron microscope image of a fabricated cavity.

Figure 1(b) shows a schematic of the sample mount used to perform strain tuning. We utilize the approach first demonstrated by Luxmoore et. al.[35] for reversible strain tuning of a photonic crystal cavity mode. An L-shaped copper holder is mounted on the cold finger of a continuous flow liquid helium cryostat. The sample and a piezo-electric actuator are mounted in parallel on the holder, with the direction of the applied stress (*y*-axis shown in the figure) aligned along the row defect of the photonic crystal cavities. The actuator is made of a 530 μm thick [Pb(Mg$_{1/3}$Nb$_{2/3}$)O$_3$]$_{0.68}$-[PbTiO$_3$]$_{0.32}$ (PMN-PT) substrate with gold coat on both the top and bottom surface to create an electrical contact. The PMN-PT substrate is poled in the [011] direction such that an out-of-plane electric field induces an anisotropic in-plane strain in the substrate. We use the convention that a positive electric field results in expansion (contraction) in the y-direction (x-direction). The device and mounting stage are cooled down to a temperature below 40 K for photoluminescence measurements.

We characterize the photoluminescence of the device by exciting the sample with a continuous wave Ti:sapphire laser tuned to a wavelength of 780 nm. The emitted light is collected by a confocal microscope using an objective lens with a numerical aperture of 0.7, and measured using a grating spectrometer with a resolution of 0.02 nm. Figure 1(c) shows the emission spectrum from a photonic crystal cavity. The spectrum exhibits a bright peak at the cavity resonant frequency that is fit to a Lorentzian function to determine a cavity quality factor of 12,000 (corresponding to a cavity decay rate of $\gamma_c/2\pi$ = 27.3 GHz). Figure 1(d) shows a temperature scan of the photoluminescence spectrum where a fixed electric field of 11.25 kV/cm was applied to the piezo during the temperature scan. The spectrum shows the cavity resonance along with multiple quantum dot lines. One of these lines, labeled as QD in the figure, anti-crosses with the cavity, indicating strong coupling. At the resonance condition attained at a temperature of 28.7 K, we perform a double-Lorentzian fit to the spectrum and measure a normal mode splitting of ΔE = 76.7 μeV. From this value we calculate the coupling strength $g$ between the quantum dot and the cavity using the relation

$$g = \sqrt{(\Delta E/2\hbar)^2 + \gamma_c^2/16}, \tag{1}$$

Plugging the previously measured value for the cavity decay rate into the above equation, we determine that $g/2\pi$ = 11.5 GHz, which satisfies the condition $g > \gamma_c/4$ ensuring that the quantum dot and the cavity are strongly coupled.[1-3]

To demonstrate strain tuning, we perform photoluminescence measurement on the same device at a fixed temperature of 30 K while sweeping the electric field applied to the piezo from zero to 15 kV/cm and then back to zero. Figure 2(a) shows the measured spectrum as a function of applied electric field. As the electric field increases, the quantum dot red-shifts across the cavity resonance, and becomes resonant at an applied field of 7.8 kV/cm. We observe a clear anti-crossing as the quantum dot exciton energy is tuned through the cavity resonance due to strong coupling. The quantum dot exciton energy returns to its original value when the field is reduced back to zero, indicating that the strain tuning process is reversible.

The redshift of the quantum dot suggests a tensile strain distribution in the center of the cavity, which is opposite to the expected direction because we are applying compressive stress. For comparison, Fig. 2(b) plots the photoluminescence as a function of electric field in the bulk region of the sample, far away from all fabricated devices. Multiple quantum dot lines appear in the measured spectrum. In contrast to the photonic crystal membrane, all bulk quantum dots experience a blueshift with increasing electric field, consistent with previous studies of quantum dots under compressive stress.[21,22] Thus, there is a large difference in the strain distribution inside the cavity membrane as compared to the bulk material.

To gain a further insight into the strain distribution inside the cavity membrane, we measure the wavelength shift for 41 different quantum dots at the center of 5 different cavities using an electric field of 15 kV/cm and plot the histogram in Fig. 2(c). All quantum dots exhibit a redshift, with an average value of $\Delta\lambda_{cav}$ = 0.45 nm, equal to 6 cavity linewidths, but there is large variations in the shift. This variation suggests that the membrane has a non-uniform strain distribution, with quantum dots at different positions experiencing different amounts of strain. In contrast, Fig. 2(d) plots a histogram of the wavelength shift at the maximum electric field for 23 individual quantum dots in the bulk far away from all cavities. The histogram shows a very narrow distribution indicating that the bulk exhibits compressive strain with a relatively uniform strain distribution. We determine a maximum shift of $\Delta\lambda_{bulk}$ = -0.52 nm (averaged over all the quantum dots) at the maximum electric field. We speculate that the unexpected strain direction in the cavity may be due to bowing or warping of the membrane, which may depend on complex factors such as size of undercut and shape of the photonic crystal.

The strain dependent spectrum in Fig. 2(a) shows that, in addition to tuning of the quantum dots, the cavity mode also exhibits a small redshift. The direction of cavity mode shift suggests a tensile strain inside the cavity membrane[34], consistent with the shift of the quantum dots. Because the strain induced cavity shift is in the same direction as the quantum dot shift, it reduces the effective tuning range of the coupled quantum dot and cavity system. Figure 3(a) shows the photoluminescence spectrum of a bare cavity mode as a function of electric field at 30 K (these measurements were performed on a different device). By fitting the emission spectrum at an electric field of zero and 15 kV/cm to a Lorentzian function (see Fig. 3(b)), we determine a spectral redshift of 0.078nm. This shift is 5.8 times smaller than the average quantum dot shift in the cavity, indicating that the cavity shift is a small effect in this system.

In conclusion, we demonstrated *in-situ* strain tuning of a quantum dot strongly coupled to a photonic crystal cavity. The average tuning range for quantum dots within the cavity is about 6 cavity linewidths, which could be further extended by applying higher electric field or utilizing multilayer piezo-electric actuators working in the longitudinal extension mode. The current device implementation can be adapted to apply strain locally to individual cavities using micro-fabricated piezoelectric micro-electro-mechanical structures.[37] The combination of local strain tuning of quantum dots with photonic crystals provides a promising path towards integrated nonlinear photonic devices for ultra-low power opto-electronics and quantum information processing on a chip.

**Acknowledgements**

The authors would like to acknowledge H. C. Materials Corporation for providing piezo samples. This work is supported by a DARPA Defense Science Office Grant (No. W31P4Q0910013), the ARO MURI on Hybrid quantum interactions Grant (No. W911NF09104), the physics frontier center at the joint quantum institute, and the ONR applied electromagnetic center N00014-09-1-1190. E. Waks would like to acknowledge support from an NSF CAREER award Grant (No. ECCS-0846494).

**Figures**

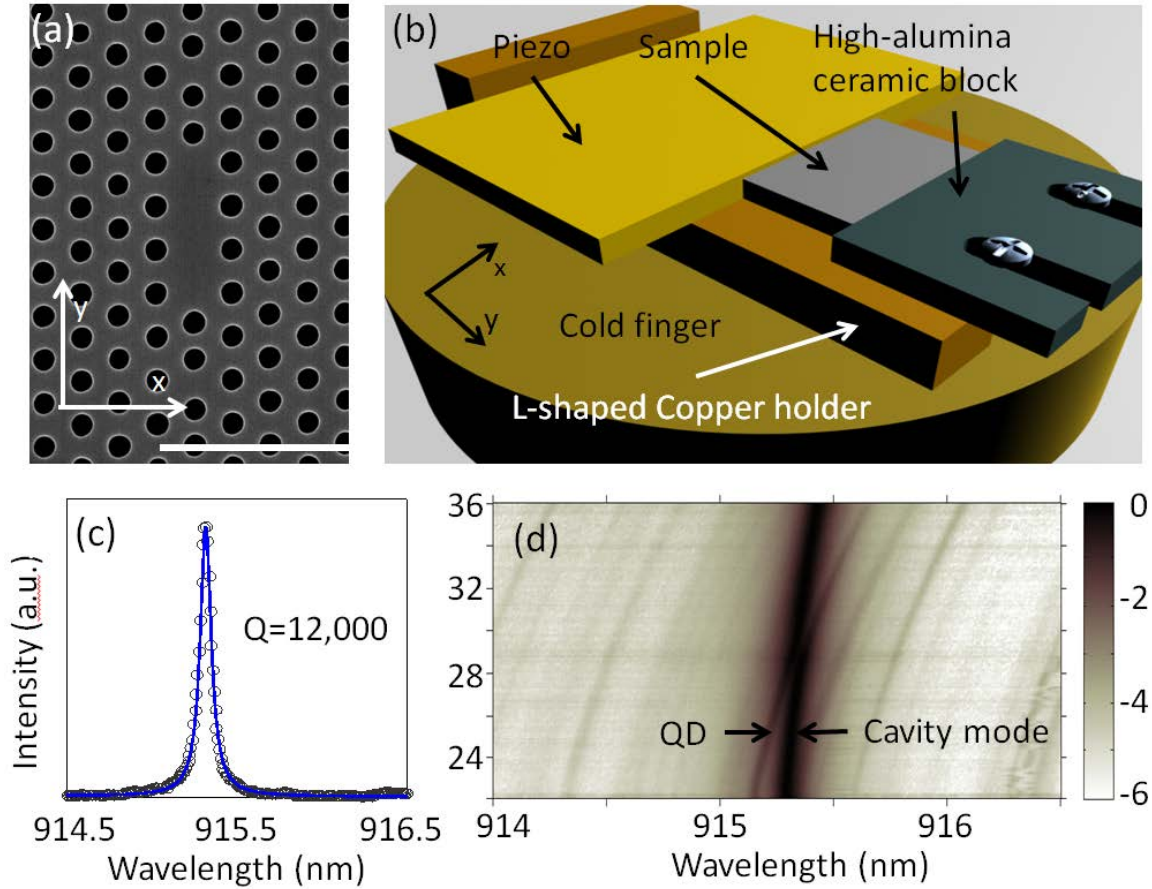

FIG. 1. (Color) (a) Scanning electron microscope image of a photonic crystal cavity. Scale bar: 1 μm. (b) Schematic diagram of the experimental setup used to apply strain to the sample. (c) Photoluminescence spectrum of a photonic crystal cavity at 30K showing a quality factor of 12,000. The black circles correspond to measured spectrum and the blue solid line shows Lorentzian fit. (d) Photoluminescence spectrum of strongly coupled quantum dot and cavity as a function of temperature.

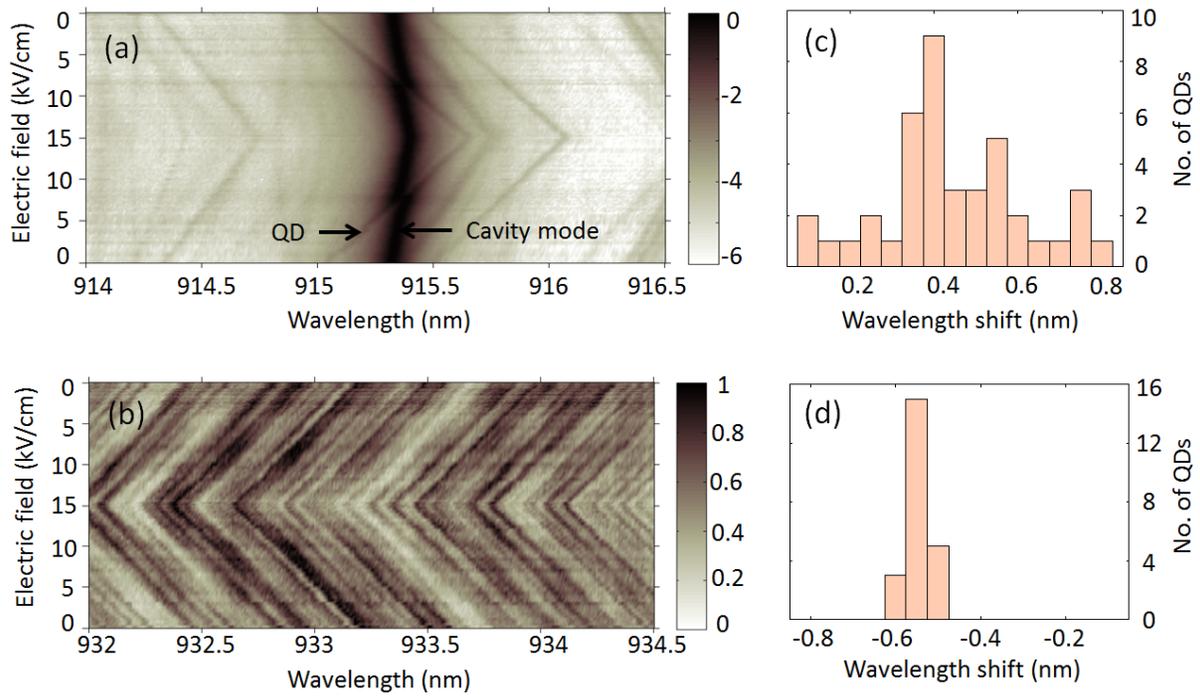

FIG. 2. (Color) (a) Photoluminescence of the cavity shown in Fig. 1(d) as a function of electric field. (b) Photoluminescence of bulk wafer containing several quantum dots as a function of electric field. (c) Histogram of the wavelength shift for 41 different quantum dots embedded in 5 different photonic crystal cavities for an electric field of 15 kV/cm. (d) Histogram of the wavelength shift for 23 different bulk quantum dots using an electric field of 15 kV/cm. The sample temperature is fixed at 30K for all above measurements.

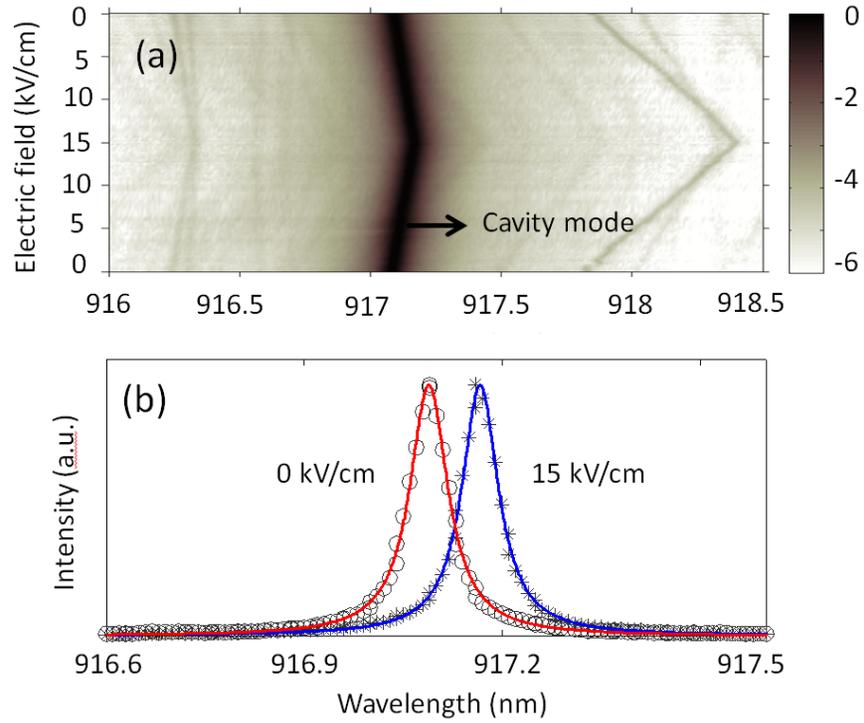

FIG. 3. (Color) (a) Photoluminescence spectrum of a cavity mode as a function of applied electric field. (b) Measured cavity spectrum at an applied electric field of zero (black circles) and 15 kV/cm (black stars). The red (blue) solid line shows Lorentzian fit to the spectrum at an electric field of zero (15 kV/cm).